# Djerfisherite: Nebular Source of Refractory Potassium

Denton S. Ebel[1,2,3], Richard O. Sack[3]

[1]American Museum of Natural History, New York, NY USA

[2]Columbia University, New York, NY USA

[3]OFM Research, Redmond, WA, USA

**Abstract:** Djerfisherite is an important carrier of potassium in highly reduced enstatite chondrites, where it occurs in sub-round metal-sulfide nodules. These nodules were once free-floating objects in the protoplanetary nebula. Here, we analyze existing and new data to derive an equation of state (EOS) for djerfisherites of $K_6(Cu,Fe,Ni)^B(Fe,Ni,Cu)_{24}^C S_{26}Cl$ structural formula. We use this EOS to calculate the thermal stability of djerfisherite coexisting in equilibrium with a cooling vapor of solar composition enriched in a dust analogous to anhydrous, chondritic interplanetary dust (C-IDP). We find that condensed mineral assemblages closely match those found in enstatite chondrites, with djerfisherite condensing above 1000 K in C-IDP dust enriched systems. Results may have implications for the volatile budgets of terrestrial planets, and the incorporation of K into early-formed, highly reduced, planetary cores. Previous work links enstatite chondrites to the planet Mercury, where the surface has a terrestrial K/Th ratio, high S/Si ratio, and very low FeO content. Mercury's accretion history may yield insights into Earth's.

**Introduction**

Djerfisherite is an accessory mineral in ultra-alkalic terrestrial magmas (e.g., Henderson et al. 1999; Zaccarini et al. 2007) and kimberlites (Clarke et al. 1994; Sharygin et al. 2007), but it is more importantly a significant carrier of potassium (K) in the metal-rich, unequilibrated (EH3) enstatite chondrites (Fuchs 1966; Keil 1968; Lin and El Goresy 2002). Djerfisherite is also found in enstatite achondrites (aubrites), and extremely rarely in sulfide nodules in iron meteorites (El Goresy et al. 1971; Kracher et al. 1977). Cosmochemists have long debated the origin of EH3 djerfisherite (Lin and El Goresy 2002; Kimura 1988), the role of enstatite chondrites in the formation of the Earth (Lodders 1995; Javoy 1995), and the potential role of K-sulfides in core segregation (Urey 1955; Wasserburg et al. 1964; Lewis 1971; Goettel 1976; Sommerville and Ahrens 1980; Lodders 1995; Chabot and Drake 1999; Rama Murthy et al. 2003; van Westrenen and Rama Murthy 2006).

Enstatite chondrite silicates are dominantly Mg-rich pyroxene ($MgSiO_3$, Weisberg and Kimura 2012). Most of their Fe is in metal and abundant metal-sulfide nodules that appear to represent unmelted analogs of the ferromagnesian chondrules common in chondrites (Lehner et al. 2010). That is, the nodules were heated and cooled as free-floating near-spherical objects in near-equilibrium with an $H_2$-rich vapor, before being accreted into the parent bodies of the enstatite chondrites (Clay et al. 2012). Potential mechanisms for local heating and cooling of such objects include shocks driven by gravitational instabilities (Desch and Connolly 2002), and current sheets driven by magnetorotational instabilities (McNally et al. 2013).

The EL enstatite chondrites contain no djerfisherite, and less metal than the EH chondrites, but both are enriched in K relative to the CV and CO carbonaceous chondrites, the ordinary chondrites, and Earth's mantle (Fig. 1). El Goresy et al. (1988) published detailed petrographic descriptions of djerfisherite-bearing nodules, and the breakdown products of primary djerfisherite in the least equilibrated EH3 chondrites. Equilibrated EH chondrites (EH4, 5) do not contain djerfisherite. The enstatite achondrites (aubrites) record melting and fractional crystallization on an enstatite chondrite-like parent body (Keil 2010; van Acken et al. 2012). Thus djerfisherite in EH3 chondrites may uniquely record conditions in reduced regions of the protoplanetary disk. Here, we investigate the nebular stability of djerfisherite as a primary condensate in the reduced vapors that may have been present during formation of enstatite chondrites,the parent material of aubrites, and the precursor material of Mercury.

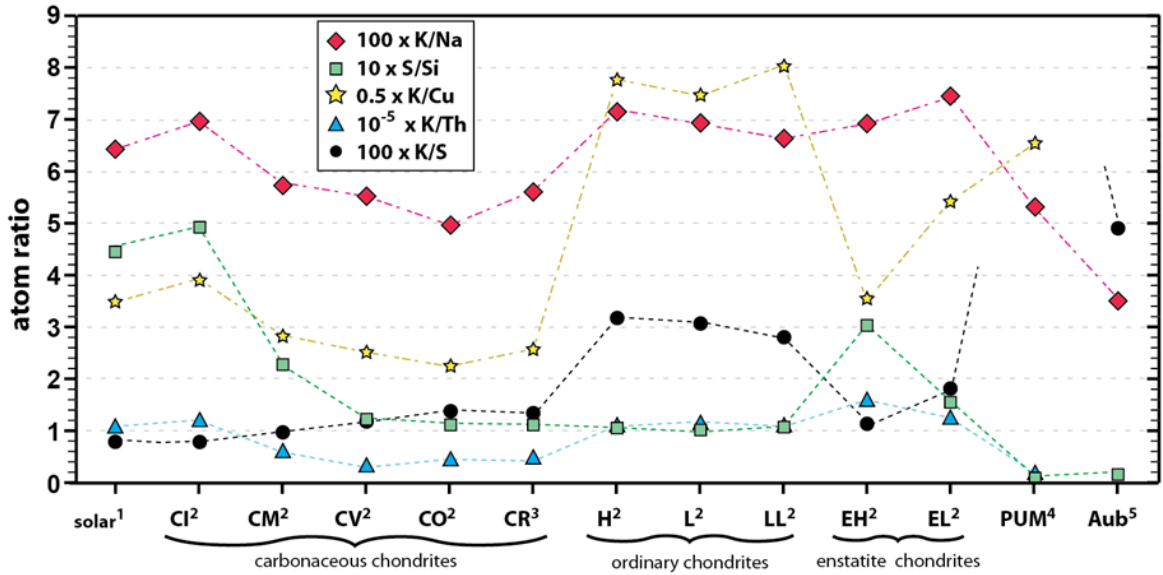

**Figure 1:** Chondrite types and element atomic abundance ratios. Data (superscripts) are from (1) Lodders (2003), (2) Wasson and Kallemeyn (1988), (3) Lodders and Fegley (1998), (4) McDonough and Sun (1995), and (5) Keil (2010), after Watters and Prinz (1979). Symbols represent: diamonds 100 K/Na, squares 10 Si/Si, stars 0.5 K/Cu, triangles $10^{-5}$ K/Th, circles 100 K/S (K/S=0.79 for PUM).

Recently, gamma and x-ray spectrometers on the MESSENGER spacecraft in orbit around Mercury returned data indicating both terrestrial K/Th (Peplowski et al. 2011) and high S/Si ratios (Nittler et al. 2011) in the Mercurian mantle. Many formation scenarios for Mercury would seem to rule out such high abundances of these volatile elements (e.g., Weidenschilling 1978; Benz et al. 1988, 2007; Fegley and Cameron 1987; Goettel 1988). Ebel and Alexander (2011) showed that sulfur behaves as a refractory element in a vapor of solar composition that is enriched in a dust of anhydrous chondritic interplanetary dust particle (C-IDP) composition (Rietmeijer 2002). A hot inner disk, well inside the snow line where water ice is lost to the solid component, may become enriched in such a dust, which has a high C/O ratio. This scenario allows for accretion of Mercury from primordial disk solids in equilibrium with vapor at high temperatures, and suggests a similar origin for the highly reduced enstatite chondrites. Potassium, however, should be strongly depleted in such a scenario. But current equilibrium condensation calculations (e.g., Ebel 2006) do not include the refractory phase djerfisherite, $\sim K_6(Fe,Ni,Cu)_{25}S_{26}Cl$, which is an important carrier of K in EH enstatite chondrites (El Goresy et al. 1988; Keil 1989). While aubrites may contain djerfisherite, it is rare (Keil 2010). Watters and Prinz (1979) used modal abundances and compositions of silicates, kamacite and troilite to calculate aubrite bulk compositions, yielding a bulk K/Na ratio of 0.037, based primarily on plagioclase (their K/Na = 0.036).

Ebel and Alexander (2011) found that CaS, FeS and MgS are stable solids above 1100 K in vapor enriched above $\sim700x$ (atoms) in C-IDP dust. Little is known about the thermodynamic properties of djerfisherite. To first order, the Gibbs energy of djerfisherite per S atom may not differ substantially from that of FeS. If so, the abundance of K in enstatite chondrites, aubrites, and Mercury may be consistent with their accretion from solids formed in equilibrium with highly reduced, C-IDP dust-enriched vapor at high temperature. Here, we estimate the thermodynamic properties of djerfisherite, calculate its stability relative to a cooling vapor of solar composition enriched in C-IDP dust, and compare the results to chemical and textural evidence from unequilibrated EH3 chondrites.

## Mineralogy

Djerfisherite is one of several chalcogenides (e.g., bartonite, pentlandite) containing $(Fe,Ni,Cu)_8S_{14}$ metal clusters (e.g., Rajamani and Prewitt 1973; Dmitrievna and Hyukhin 1975; Evans and Clark 1981). It is cubic, with space group symmetry *Pm3m*, and typically approximates the structural formula $(K,Na)_6^A (Na,Li,Cu,Fe,Ni)^B (Fe,Ni,Cu)_{24}^C S_{26}Cl$, with the metals in B and C sites in octahedral and tetrahedral coordination, respectively. Variable degrees of vacancies are reported for these, and for the A and Cl sites (e.g., Clarke et al. 1994; Barkovi et al. 1997; Sharygin et al. 2007; Zaccarini et al. 2007). Na, Li, and Cu occupancy of the B site has been confirmed for natural and synthetic djerfisherites (e.g., Czamanske et al. 1979; Tani 1977; Zaccarini et al. 2007). Occupancy of the B site by Fe, and possibly also Ni, appears to be required by the stability of djerfisherites synthesized by Clarke (1979) in the simple systems K-Fe-S-Cl and K-Fe-Ni-S-Cl and by djerfisherites in EH chondrites (Fuchs 1966; El Goresy et al. 1988; Lin and El Goresy 2002). The synthetic djerfisherites have Fe/S and (Fe+Ni)/S ratios within analytical uncertainly of a formula based on 25 metal and 26 sulfur atoms [Fe = $25.285 \pm 0.050$ and (Fe+Ni) = $25.580 \pm 0.524$ per 26 atom S formula unit]. Analyses (n=77) of djerfisherites from Qingzhen (EH3; Lin, El Goresy, pers. comm.) have sums of metals indistinguishable from 25 ($25.077 \pm 0.390$) and have $2 \geq$ Cu $\geq 1$ ($1.322 \pm 0.337$) on a 26 S formula basis. By extension, these analyses suggest that Na occupies the A site, as (K + Na) $\leq 6$ ($5.820 \pm 0.053$) and Na $> 0.5$ ($0.718 \pm 0.129$) on this basis (Lin and El Goresy 2002).

For the purpose of our calculations we consider that djerfisherites approximate the structural formula $(K,Na)_6^A (Cu,Fe,Ni)^B (Fe,Ni,Cu)_{24}^C S_{26}Cl$. We will describe djerfisherite using three independent composition variables: $X_2$ = Ni/(Cu+Ni+Fe), $X_3$ = Cu/(Cu+Ni+Fe) and $X_4$ = Na/(Na+K) with the average djerfisherite in Qingzhen having values of $X_2 = 0.0264$, $X_3 = 0.0527$, and $X_4 = 0.1322$. However, because we cannot constrain the K-Na exchange properties with existing data, we consider Na-K exchange to be ideal and restrict calculations and discussion to $K_6(Cu,Fe,Ni)^B (Fe,Ni,Cu)_{24}^C S_{26}Cl$.

## Djerfisherite Thermodynamic Properties

The synthesis experiments of Clarke (1979) constrain the value of the Gibbs energy of formation of djerfisherite (djr) from Fe metal (Fe), troilite (tro, FeS), potassium thioferrite (kfs, $KFeS_2$), and KCl in the K-Fe-S-Cl system to zero at T = $324 \pm 28°C$ and 1 atmosphere pressure. It seems warranted to assume that kinetic factors did not prevent djerfisherite formation at the lowest temperature investigated by Clarke (296°C, 192 hr), based on the authors' experience in arguably more refractory sulfide systems at temperatures down to 200°C (e.g., Ebel and Sack 1994). In a simple approximation, we assume that the higher temperature stability of djerfisherite in the reaction

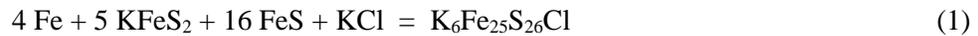

$$4\ Fe + 5\ KFeS_2 + 16\ FeS + KCl\ =\ K_6Fe_{25}S_{26}Cl \tag{1}$$

is due entirely to the greater configuration entropy of djerfisherite relative to the reactants in (1) and that the quantity $\Delta \overline{S}_1^o$ is temperature independent. To first order, we may determine this entropy from the two reactions whose sum is reaction (1):

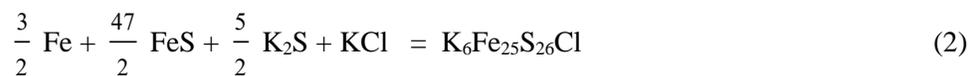

$$\frac{3}{2}\ Fe + \frac{47}{2}\ FeS + \frac{5}{2}\ K_2S + KCl\ =\ K_6Fe_{25}S_{26}Cl \tag{2}$$

and
$$\frac{5}{2}\,\text{Fe} + 5\,\text{KFeS}_2 \; = \; \frac{5}{2}\,\text{K}_2\text{S} + \frac{15}{2}\,\text{FeS}\;. \qquad (3)$$

We may estimate the entropy of the first of these reactions, $\Delta \bar{S}_2^o$, using the formalism of Craig and Barton (1973):

$$\Delta \bar{S}_2^o = - (1.2 \pm 0.8)\,\text{R} \sum_i N_i \ln N_i \qquad (4)$$

where $N_i$ are the mole fractions of simple sulfides with one sulfur atom in their formula unit and $\Delta \bar{S}^o$ is the entropy of formation of the chalcogenide from the simple sulfides per gram atom of sulfur, a formalism which produces estimates of 82.11 ± 54.74 JK$^{-1}$-mol$^{-1}$ for $\Delta \bar{S}_2^o$ (or a virtually identical entropy of 80.53 ± 53.69 if the mole fractions of simple sulfides are computed using all species in reaction 2). The entropy of reaction (3) can readily be computed to be 51.75 J/K$^{-1}$-mol$^{-1}$ based on entropy data from the JANAF database (Chase 1998), and from the 298.15°C estimate for the entropy of potassium thioferrite, $\Delta \bar{S}_{298,kfs}^o =$ 124.0 JK$^{-1}$mol$^{-1}$ given by Latimer's rules (Sommerville and Ahrens, 1980). Summing the entropies for reactions (2) and (3), we obtain 133.87 ± 54.74 JK$^{-1}$mol$^{-1}$ for the entropy of reaction (1), and 79.940 ± 32.688 kJmol$^{-1}$ for $\Delta \bar{H}_1^o$, using $\Delta \bar{H}_1^o =$ T$\Delta \bar{S}_1^o$ at 324°C. Finally, using Sommerville and Ahrens' (1980, their Table 5) estimate for the Gibbs energy of formation of potassium thioferrite (*kfs*), and the entropy of formation of KFeS$_2$, we obtain $\Delta \bar{H}_{fm,kfs}^o = \Delta \bar{G}_{fm,kfs}^o + T\Delta \bar{S}_{fm,kfs}^o = -350.471$ kJmol$^{-1}$. Then $\Delta \bar{H}_3^o$ = 48.425 kJmol$^{-1}$, $\Delta \bar{G}_3^o$ = 32.990 kJmol$^{-1}$, and $\Delta \bar{S}_3^o$ = 51.75 JK$^{-1}$mol$^{-1}$. These results permit us, finally, to derive $\Delta \bar{H}_2^o$ = 31.515 kJ/mol$^{-1}$ at 298 K.

Combining results for reaction (1), $\Delta \bar{H}_1^o =$79.940 kJmol$^{-1}$, $\Delta \bar{S}_1^o =$133.87 JK$^{-1}$mol$^{-1}$, $\Delta \bar{G}_1^o = \Delta \bar{H}_1^o$ -T$\Delta \bar{S}_1^o =$ 40.027 kJmol$^{-1}$, with results for $\Delta \bar{G}_{kfs}^o$, and $\Delta \bar{G}_{formation}^o$ for Fe, FeS, and KCl from the JANAF tables, the Gibbs energy of formation of djerfisherite K$_6$Fe$_{25}$S$_{26}$Cl can be computed at 298 K as -3690.81 kJmol$^{-1}$. The temperature dependence of all these properties is not known. Assuming constant values for $\Delta \bar{H}_1^o =$79.940 kJmol$^{-1}$, and $\Delta \bar{S}_1^o =$133.87 JK$^{-1}$mol$^{-1}$, and using JANAF values for Fe, FeS, and KCl, and derived values (above) for KFeS$_2$, reaction (1) has a net negative Gibbs energy above ~595 K (Table 1).

| T(K) | $\Delta \overline{G}^o_{LHS,1}$ (kJ) | $\Delta \overline{H}^o_{LHS,1}$ (kJ) | $\Delta \overline{G}^o_1$ (J) | $\Delta \overline{G}^o_{fm,djr}$ (kJ) |
|---|---|---|---|---|
| 298.15 | -3730.84 | -3800.88 | 40027 | -3690.811 |
| 300 | -3730.69 | -3801.98 | 39779 | -3690.91 |
| 400 | -3723.80 | -3927.17 | 26392 | -3697.409 |
| 500 | -3721.67 | -4004.94 | 13005 | -3708.660 |
| 600 | -3722.19 | -4116.23 | -382 | -3722.570 |
| 700 | -3724.27 | -4261.59 | -13769 | -3738.038 |
| 800 | -3725.02 | -4421.46 | -27156 | -3752.173 |
| 900 | -3706.56 | -5989.52 | -40543 | -3747.105 |
| 1000 | -3607.54 | -6214.39 | -53930 | -3661.467 |
| 1100 | -3501.96 | -6959.73 | -67317 | -3569.273 |
| 1200 | -3392.44 | -7230.92 | -80704 | -3473.139 |
| 1300 | -3283.13 | -7467.23 | -94091 | -3377.224 |
| 1400 | -3175.69 | -7702.71 | -107478 | -3283.171 |
| 1500 | -3083.04 | -7440.20 | -120865 | -3203.904 |
| 1600 | -3013.23 | -7697.20 | -134252 | -3147.486 |

**Table 1:** Results calculated for reaction (1) assuming constant values for $\Delta \overline{S}^o_1 = 133.87$ J, and $\Delta \overline{H}^o_1 = 79940$ kJ. The LHS (left hand side, or reactants) summed $\Delta \overline{G}^o_{LHS,1}$ and $\Delta \overline{H}^o_{LHS,1}$ are from JANAF data and derived values for KFeS$_2$ (see text). Given $\Delta \overline{G}^o_1$, the Gibbs energy of formation of djerfisherite from the elements in their standard states at the temperature of interest $\Delta \overline{G}^o_{fm,djr}$ may be calculated.

**Nickel Substitution**

The Gibbs energy of formation of the endmember nickel djerfisherite, K$_6$Ni$_{24}$S$_{26}$Cl, may be constrained using Fe-Ni partitioning data involving djerfisherite and other Fe- and Ni-bearing phases (e.g., (Fe,Ni)$_{1\pm x}$S monosulfide solid solution (mss), Fe-Ni metal alloy) to determine $\Delta \overline{H}^o$ 's of Fe-Ni exchange reactions such as

$$\frac{1}{25} K_6Fe_{25}S_{26}Cl + Ni^{metal} = Fe^{metal} + \frac{1}{25} K_6Ni_{25}S_{26}Cl \qquad (5)$$

under the simplifying assumption that the $\Delta \overline{S}^o$ 's of such exchange reactions are zero (e.g., Sack and Ebel 2006). Analysis of the compositions of coexisting kamacite and djerfisherite is performed below using the following condition of equilibrium for reaction (5):

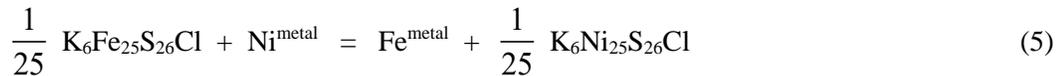

$$\frac{1}{25} \Delta \bar{G}^o_{Ni-djr} = \frac{1}{25} \Delta \bar{G}^o_{Fe-djr} + \Delta \bar{G}^o_{Ni-metal} - \Delta \bar{G}^o_{Fe-metal} - RTln\left(\frac{a_{Fe}}{a_{Ni}} \cdot \frac{a_{Ni-djr}^{1/25}}{a_{Fe-djr}^{1/25}}\right). \qquad (6)$$

**Thermodynamic Formulation**

Based on the results of structure refinements, it would appear that Cu is strongly partitioned into the B site relative to Fe and Ni, requiring the choice of K$_6$Cu$^B$Fe$^C_{24}$S$_{26}$Cl as an ordered component, and necessitating, at a minimum, that the ordering reaction

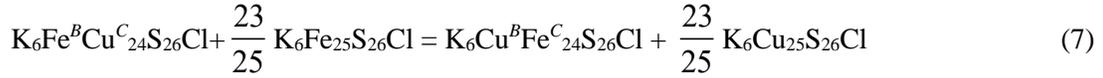

$$K_6Fe^BCu^C_{24}S_{26}Cl + \frac{23}{25}\ K_6Fe_{25}S_{26}Cl = K_6Cu^BFe^C_{24}S_{26}Cl + \frac{23}{25}\ K_6Cu_{25}S_{26}Cl \qquad (7)$$

be evaluated in addition to the Fe-Cu exchange reaction

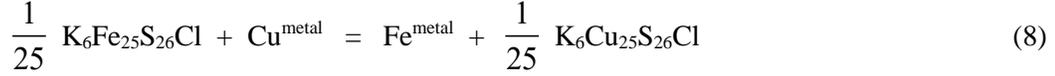

$$\frac{1}{25}\ K_6Fe_{25}S_{26}Cl\ +\ Cu^{metal}\ =\ Fe^{metal}\ +\ \frac{1}{25}\ K_6Cu_{25}S_{26}Cl \qquad (8)$$

and the Fe-Cu reciprocal-ordering reaction

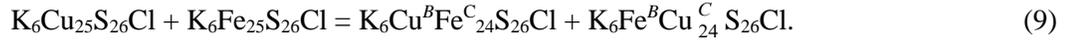

$$K_6Cu_{25}S_{26}Cl + K_6Fe_{25}S_{26}Cl = K_6Cu^BFe^C_{24}S_{26}Cl + K_6Fe^BCu^C_{24}S_{26}Cl. \qquad (9)$$

In a first approximation, we may assume that the Gibbs energy of reciprocal reaction (9) and the entropies of the ordering reaction (7) and the Fe-Cu exchange reaction (8) are zero. For these assumptions, only the enthalpies of the Fe-Cu B-C site ordering reaction (7) and Fe-Cu exchange reaction (8) need be evaluated, and we may readily develop the following equation for the Gibbs energy of $K_6(Cu,Fe,Ni)^B$ $(Fe,Ni,Cu)^C_{24}S_{26}Cl$ djerfisherite:

$$\overline{G} = \overline{G}^o_1(1 - X_2 - X_3) + \overline{G}^o_2\,X_2\ + \overline{G}^o_3\,X_3 + \frac{1}{2}\Delta\overline{G}^{o-ORD}_{Ni-Fe}(s_1) + \frac{1}{2}\Delta\overline{G}^{o-ORD}_{Cu-Fe}(s_2)$$

$$+\ RT\ [(1 - X_2 - X_3 - \frac{24}{25}\,(s_1 + s_2))\ \ln\,(1 - X_2 - X_3 - \frac{24}{25}\,(s_1 + s_2))$$

$$+ (X_2 + \frac{24}{25}\,s_1)\ \ln\,(X_2 + \frac{24}{25}\,s_1) + (X_3 + \frac{24}{25}\,s_2)\ \ln\,(X_3 + \frac{24}{25}\,s_2)$$

$$+ 24\,(1 - X_2 - X_3 + \frac{1}{25}\,(s_1 + s_2))\ \ln\,(1 - X_2 - X_3 + \frac{1}{25}\,(s_1 + s_2))$$

$$+ 24\,(X_2 - \frac{1}{25}\,s_1)\ \ln\,(X_2 - \frac{1}{25}\,s_1) + 24\,(X_3 - \frac{1}{25}\,s_2)\ \ln\,(X_3 - \frac{1}{25}\,s_2)] \qquad (10)$$

when we define $X_2 \equiv Ni/(Fe+Ni+Cu)$, $X_3 \equiv Cu/(Fe+Ni+Cu)$, $s_1 \equiv X^B_{Ni}$ - $X^C_{Ni}$ and $s_2 \equiv X^B_{Cu}$ - $X^C_{Cu}$ as composition and ordering variables, and assume the vibrational Gibbs energy may be described by a Taylor expansion of first degree in these variables (e.g., Thompson 1969). In this equation the terms $\overline{G}^o_1$, $\overline{G}^o_2$, $\overline{G}^o_3$ are the Gibbs energy of formation of the djerfisherite endmembers $K_6Fe_{25}S_{26}Cl$, $K_6Ni_{25}S_{26}Cl$ and $K_6Cu_{25}S_{26}Cl$, $\Delta\overline{G}^{o-ORD}_{Ni-Fe}$ and $\Delta\overline{G}^{o-ORD}_{Cu-Fe}$ are the Gibbs energy of the ordering reaction

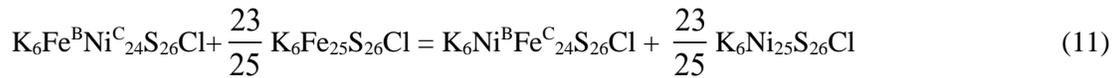

$$K_6Fe^BNi^C_{24}S_{26}Cl + \frac{23}{25}\ K_6Fe_{25}S_{26}Cl = K_6Ni^BFe^C_{24}S_{26}Cl + \frac{23}{25}\ K_6Ni_{25}S_{26}Cl \qquad (11)$$

and reaction (7), and where we may identify atom site fractions with these composition and ordering variables

$$X^B_{Fe} = 1 - X_2 - X_3 - \frac{24}{25}\,(s_1 + s_2)$$

$$X^B_{Ni} = X_2 + \frac{24}{25}\,s_1$$

$$X^B_{Cu} = X_3 + \frac{24}{25}\,s_2$$

$$X_{Fe}^{C} = 1 - X_2 - X_3 + \frac{1}{25}(s_1 + s_2) \tag{12}$$

$$X_{Ni}^{C} = X_2 - \frac{1}{25} s_1$$

$$X_{Cu}^{C} = X_3 - \frac{1}{25} s_{2.}$$

Finally, the ordering variables $s_1$ and $s_2$ may be evaluated by setting the first partial derivatives of the Gibbs energy with respect to these variables equal to zero, a procedure (see Appendix I) which results in the expressions

$$RT \ln [(X_{Fe}^{B} X_{Ni}^{C})/(X_{Fe}^{C} X_{Ni}^{B})] = \frac{25}{48}\Delta \bar{G}_{Ni-Fe}^{o-ORD}$$

$$\tag{13}$$

$$RT \ln [(X_{Fe}^{B} X_{Cu}^{C})/(X_{Fe}^{C} X_{Cu}^{B})] = \frac{25}{48}\Delta \bar{G}_{Cu-Fe}^{o-ORD},$$

The chemical potentials of endmember components $j$ may be evaluated from the familiar expression

$$\mu_j = \bar{G} + \sum_i n_{ij}(1-X_i)\left(\frac{\partial \bar{G}}{\partial X_i}\right)_{X_k/X_1} + (q_{1j}-s_1)\left(\frac{\partial \bar{G}}{\partial s_1}\right)_{T,X_2,X_3,s_2} + (q_{2j}-s_2)\left(\frac{\partial \bar{G}}{\partial s_2}\right)_{T,X_2,X_3,s_1} \tag{14}$$

where the sum is over all linearly independent compositional components $i$, $n_{ij}$ is the stoichiometric coefficient of component $i$ per mole of $j$, and $q_{zj}$ is the value of the order variable $z$ in component $j$ (cf. Sack et al. 1987). Utilizing (14) we readily obtain

$$\mu_1 = \bar{G}_1^{o} + RT \ln (X_{Fe}^{B})(X_{Fe}^{C})^{24}$$

$$\mu_2 = \bar{G}_2^{o} + RT \ln (X_{Ni}^{B})(X_{Ni}^{C})^{24} \tag{15}$$

$$\mu_3 = \bar{G}_3^{o} + RT \ln (X_{Cu}^{B})(X_{Fe}^{C})^{24}$$

Even in the absence of any experimental constraints, we may reasonably assume that the enthalpy of reaction (7) is strongly negative, as noted above, and assume that there is no ordering of Fe and Ni between B and C sites (i.e., set $\Delta \bar{G}_{Ni-Fe}^{o-ORD}=0$). For these, and previous, assumptions, we may readily conclude that $\Delta \bar{G}_{Cu-Fe}^{o-ORD} \leq -70$ kJ to guarantee that $X_{Fe}^{B} \geq 0.95$ in $K_6(Fe,Cu)_{25}S_{26}Cl$ djerfisherites with one or more atoms of Cu per formula unit (pfu) at 200°C. This approximate upper bound is predicated on the observations that (1) only Li, Na and Cu have been proposed as occupants of the B site based on the results of structural studies (e.g., Czamanske et al. 1979; Tani 1977; Evans and Clark 1981) and (2) Fe appears to readily enter the B site in the absence of Cu, Na and Li (e.g., Clarke 1979; Zaccarini et al. 2007). More negative values of $\Delta \bar{G}_{Cu-Fe}^{o-ORD}$ appear to be required (e.g., < − 130 kJ, Fig. 2) to stabilize the compositions of djerfisherites observed in EH chondrites. They further exclude Fe from the B site, progressively lowering the activity of the $K_6Fe_{25}S_{26}Cl$ component relative to an ideal, disordered, multisite mixture (Fig. 3), and stabilize chondritic djerfisherites with respect to reaction (1) down to below around 273 K according to the relation

$$T = \Delta \bar{H}_1^{o}/(\Delta \bar{S}_1^{o} - R \ln a_{K_6Fe_{25}Cu_{26}Cl}), \tag{16}$$

obtained from reaction (1) when only solid solution effects in djerfisherite are considered. In contrast to activity-composition relations for $K_6Fe_{25}S_{26}Cl$, activity-composition relations are nearly ideal for the $K_6Cu_{25}S_{26}Cl$ component, except in very Cu-poor compositions (Fig. 3) exemplified by those found in the Qingzhen meteorite, where they exhibit pronounce negative deviations from ideality.

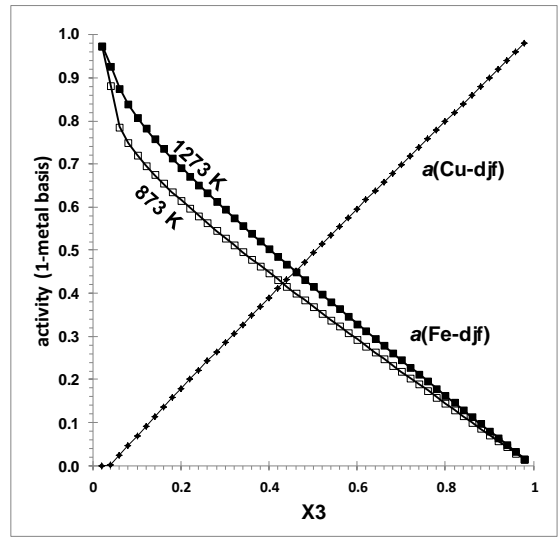

**Figure 3:** Activity and composition relationships calculated for $K_{6/25}(Fe,Cu)S_{26/25}Cl_{1/25}$ djerfisherites, $X_3 = Cu/(Fe+Ni+Cu)$, assuming $\Delta \bar{G}_{Cu-Fe}^{o-ORD}$ = -130 kJ at 873 and 1273 K.

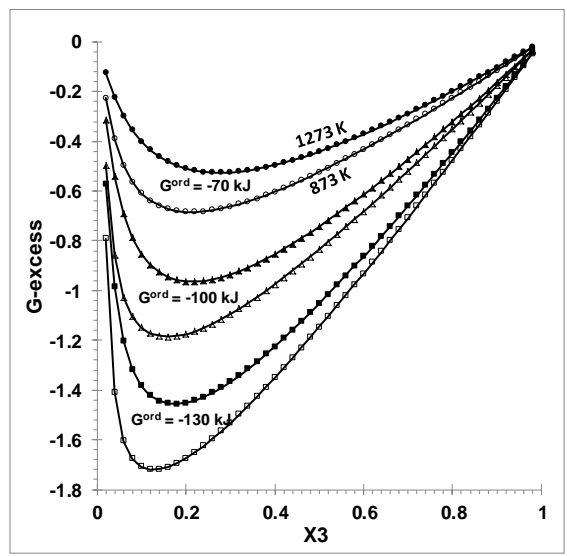

**Figure 2**: Excess Gibbs energy of mixing (kJ) of Cu-Fe djerfisherite solutions expressed on a one Cu-Fe formula basis (i.e. $K_{6/25}(Fe,Cu)S_{26/25}Cl_{1/25}$). Open symbols are for 873 K, filled symbols for 1273 K, for $\Delta \bar{G}_{Cu-Fe}^{o-ORD}$ = -70 (circles), -100 (triangles) and -130 kJ/gfw (squares).

**Petrographic Constraints**

A large djerfisherite clast in EH3 chondrite Sahara 97096 (AMNH 4940-1) was analyzed for the compositions of coexisting phases, using the SX100 electron probe micro analyzer (EPMA) at AMNH. Methods are outlined in Appendix II. Low totals for alkali elements reflect a lack of appropriate standards, but should not affect analysis of Fe-Ni partitioning with metal. The compositions of coexisting djerfisherite and metal alloy (Table 2), were used to constrain the Gibbs energy of the Ni-endmember $K_6Ni_{25}S_{26}Cl$, $\overline{G}_2^o$, using equation (6), where the activities of Fe and Ni in metal alloy are calculated as in Ebel and Grossman (2000), and the activities of djerfisherite components $a_{Fe\text{-}djr}$ and $a_{Ni\text{-}djr}$ are set to the mean values of their mole fractions, $X_{Fe\text{-}djr} = 0.8885$ and $X_{Ni\text{-}djr} = X_2 = 0.0254$ (Table 2). A similar calculation yielded $\Delta \overline{G}_3^o$, with $X_{Cu\text{-}djr} = X_3 = 0.0858$, setting $a_{Cu}$ in metal to its mole fraction.

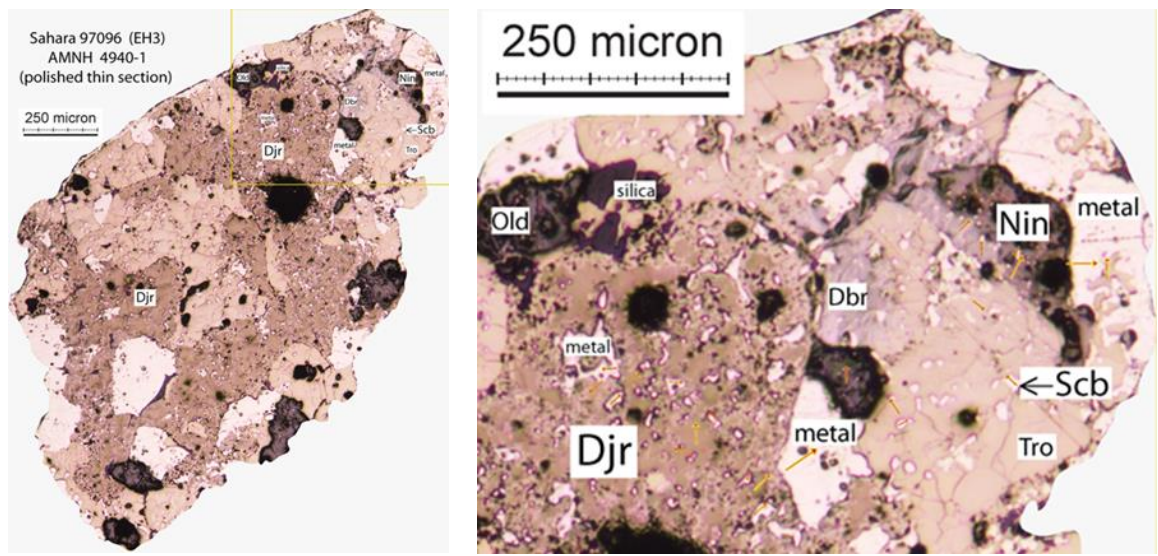

**Figure 4:** Large djerfisherite-rich clast in EH3 chondrite SAH 97096 (AMNH section 4940-1). Reflected light image of entire clast is at left, and detail (right) shows EPMA analysis lines. Phases include: djerfisherite (Djr), schreibersite (Scb), niningerite (Nin), daubreelite (Dbr), oldhamite (Old), troilite (Tro), graphite (dark, round), silica, and metal. Large djerfisherite grains contain numerous metal blebs, and have edges altered to secondary troilite in a 'Qingzhen reaction' texture (El Goresy et al. 1988).

| | Na | Mg | Al | Si | K | Cl | P | S | Ca | Ti | Cr | Mn | Fe | Co | Ni | Cu | Zn | Total | n |
|---|---|---|---|---|---|---|---|---|---|---|---|---|---|---|---|---|---|---|---|
| **Metal** | 0.014 | 0.092 | 0.037 | 2.617 | 0.005 | 0.008 | 0.030 | 0.077 | 0.001 | 0.006 | 0.025 | 0.049 | 92.353 | 0.384 | 3.759 | 0.011 | 0.088 | 99.56 | 36 |
| stdev | 0.009 | 0.311 | 0.044 | 0.180 | 0.006 | 0.009 | 0.021 | 0.165 | 0.002 | 0.008 | 0.054 | 0.113 | 1.437 | 0.025 | 0.866 | 0.021 | 0.022 | 0.64 | |
| formula | 0.000 | 0.002 | 0.001 | 0.051 | 0.000 | 0.000 | 0.000 | | 0.000 | 0.000 | 0.000 | 0.000 | 0.904 | 0.004 | 0.035 | 0.000 | 0.001 | **1.0** | |
| | | | | | | | | | | | | | | | | | | | |
| **Djerfisherite** | 0.690 | 0.009 | 0.004 | 0.040 | 6.960 | 1.440 | 0.010 | 33.564 | 0.004 | 0.003 | 0.008 | 0.015 | 49.949 | 0.007 | 1.502 | 5.486 | 0.246 | 99.94 | 30 |
| stdev | 0.187 | 0.007 | 0.007 | 0.013 | 0.781 | 0.105 | 0.011 | 0.440 | 0.004 | 0.005 | 0.009 | 0.016 | 1.229 | 0.010 | 0.166 | 0.581 | 0.031 | 0.82 | |
| formula | 0.746 | 0.009 | 0.004 | 0.035 | 4.420 | 1.009 | 0.008 | **26.0** | 0.002 | 0.002 | 0.004 | 0.007 | 22.212 | 0.003 | 0.635 | 2.144 | 0.093 | | |
| | | | | | | | | | | | | | | | | | | | |
| **Schreibersite** | 0.003 | 0.002 | 0.002 | 0.257 | 0.012 | 0.005 | 15.149 | 0.092 | 0.001 | 0.008 | 0.129 | 0.003 | 67.321 | 0.083 | 16.597 | 0.047 | 0.097 | 99.81 | 18 |
| stdev | 0.007 | 0.002 | 0.003 | 0.022 | 0.008 | 0.006 | 0.207 | 0.220 | 0.002 | 0.011 | 0.159 | 0.006 | 1.243 | 0.024 | 1.148 | 0.035 | 0.016 | 1.17 | |
| formula | 0.000 | 0.000 | 0.000 | 0.019 | 0.001 | 0.000 | **1.0** | 0.006 | 0.000 | 0.000 | 0.005 | 0.000 | 2.465 | 0.003 | 0.578 | 0.002 | 0.003 | | |
| | | | | | | | | | | | | | | | | | | | |
| **Troilite** | 0.011 | 0.011 | 0.003 | 0.055 | 0.005 | 0.013 | 1.179 | 34.276 | 0.038 | 0.269 | 2.227 | 0.069 | 59.749 | 0.013 | 1.594 | 0.134 | 0.299 | 99.95 | 9 |
| stdev | 0.028 | 0.018 | 0.004 | 0.039 | 0.005 | 0.018 | 3.521 | 8.051 | 0.104 | 0.047 | 1.254 | 0.034 | 1.625 | 0.032 | 4.258 | 0.091 | 0.063 | 0.53 | |
| formula | 0.000 | 0.000 | 0.000 | 0.002 | 0.000 | 0.000 | 0.036 | **1.0** | 0.001 | 0.005 | 0.040 | 0.001 | 1.001 | 0.000 | 0.025 | 0.002 | 0.004 | | |
| | | | | | | | | | | | | | | | | | | | |
| **Daubreelite** | 0.063 | 0.013 | 0.010 | 0.048 | 0.009 | 0.006 | 0.007 | 43.140 | 0.007 | 0.183 | 30.742 | 0.513 | 21.877 | 0.006 | 0.103 | 0.589 | 0.604 | 97.92 | 7 |
| stdev | 0.109 | 0.003 | 0.004 | 0.009 | 0.011 | 0.009 | 0.008 | 1.045 | 0.007 | 0.143 | 5.158 | 0.057 | 7.311 | 0.010 | 0.064 | 0.218 | 0.251 | 1.12 | |
| formula | 0.008 | 0.002 | 0.001 | 0.005 | 0.001 | 0.001 | 0.001 | **4.0** | 0.000 | 0.011 | 1.758 | 0.028 | 1.164 | 0.000 | 0.005 | 0.028 | 0.027 | | |
| | | | | | | | | | | | | | | | | | | | |
| **Niningerite** | 0.063 | 25.422 | 0.011 | 0.044 | 0.003 | 0.008 | 0.008 | 48.323 | 0.316 | 0.008 | 0.280 | 11.296 | 14.613 | 0.018 | 0.085 | 0.249 | 0.134 | 100.88 | 6 |
| stdev | 0.012 | 0.309 | 0.011 | 0.011 | 0.004 | 0.011 | 0.008 | 0.482 | 0.031 | 0.010 | 0.072 | 0.689 | 1.042 | 0.009 | 0.040 | 0.033 | 0.036 | 0.49 | |
| formula | 0.002 | 0.694 | 0.000 | 0.001 | 0.000 | 0.000 | 0.000 | **1.0** | 0.005 | 0.000 | 0.004 | 0.136 | 0.174 | 0.000 | 0.001 | 0.003 | 0.001 | | |
| | | | | | | | | | | | | | | | | | | | |
| **Cr-sulfide** | 0.242 | 0.016 | 0.015 | 0.041 | 0.073 | 0.004 | 0.004 | 46.079 | 0.025 | 0.081 | 36.794 | 0.080 | 1.788 | 0.011 | 0.056 | 0.742 | 0.076 | 86.13 | 5 |
| stdev | 0.347 | 0.019 | 0.005 | 0.007 | 0.056 | 0.003 | 0.004 | 3.317 | 0.040 | 0.029 | 2.546 | 0.075 | 2.667 | 0.016 | 0.006 | 0.530 | 0.023 | 2.46 | |
| formula | 0.015 | 0.001 | 0.001 | 0.002 | 0.003 | 0.000 | 0.000 | **2.0** | 0.001 | 0.007 | 0.985 | 0.002 | 0.045 | 0.000 | 0.001 | 0.016 | 0.002 | | |
| | | | | | | | | | | | | | | | | | | | |
| **Oldhamite** | 0.017 | 0.459 | 0.001 | 0.041 | 0.000 | 0.017 | 0.000 | 42.819 | 53.650 | 0.002 | 0.000 | 0.151 | 0.572 | 0.030 | 0.090 | 0.124 | 0.068 | 98.04 | 1 |
| formula | 0.001 | 0.014 | 0.000 | 0.001 | 0.000 | 0.000 | 0.000 | **1.0** | 1.002 | 0.000 | 0.000 | 0.002 | 0.008 | 0.000 | 0.001 | 0.001 | 0.001 | | |

**Table 2** Element wt%, standard deviation, and formulae for coexisting phases in the clast of Fig. 4. Formulae for djerfisherite, troilite, daubreelite, niningerite and oldhamite are normalized to the highlighted number of atoms per unit of sulfur or phosphorus. Schreibersite is normalized to 1 phosphorus, and metal to one atom total.

## Nebular Stability Calculations

The same nebular dust compositions previously examined by Ebel and Alexander (2012) were considered here. The C-IDP dust analog is an H-, N-, F-, Cl-free dust of chondritic composition (Lodders 2003, Table 1, col 3), retaining all C as elemental (~graphitic) carbon, all S as FeS, Co and Ni as metal, and only enough O to make oxides of the remaining Fe and of all Si, Mg, Ca, Al, Na, P, K, Ti, Cr, Mn. This C-IDP dust contains most elements at chondritic ratios, but less O than the CI dust considered by Ebel and Grossman (2000). The analog dust was added to complementary solar composition vapor (enrichment of 1 = solar) in various proportions to make the bulk compositions. The 'VAPORS code of Ebel and Grossman (2000; cf., Ebel, 2006) was used to calculate thermodynamic equilibria between fully speciated vapor, silicate liquid, and solid minerals at 5 or 10° steps at fixed total pressure ($P^{tot}$) of 10Pa ($10^{-4}$ bar) for each bulk composition.

The VAPORS code uses equation of state data consistent with the 'Benson-Helgeson' convention, describing the 'apparent' Gibbs energies of compounds as referred to those of the elements in their standard states at 298 K and one bar (Anderson and Crerar 1993). The preceding analysis of published data, a value of $\Delta \bar{G}^{o-ORD}_{Cu-Fe} = -130$ kJmol$^{-1}$, and constraints from EPMA analysis of phase assemblages were combined to provide estimates of the apparent Gibbs energies of formation of djerfisherite of composition $X_2 = 0.0193$, $X_3 = 0.0631$ from 1600 K to ~1100 K, using equation 10. These were parameterized for the condensation calculation (Table 3).

| endmember | $\Delta \bar{G}^o_{ap}$ (**1600 K**) | $d\bar{G}^o_{ap}/dT$ |
|---|---|---|
| $K_6Fe_{25}S_{26}Cl$ | -1.0 x $10^7$ Jmol$^{-1}$ | -3000 JK$^{-1}$ |
| $K_6Ni_{25}S_{26}Cl$ | -9.98066 x $10^6$ | -2886 |
| $K_6Cu_{25}S_{26}Cl$ | -1.00992 x $10^7$ | -2903 |

**Table 3** Parameterization of apparent Gibbs energy of formation of djerfisherite endmembers for application between 1600 and 1100 K.

Djerfisherite stability was calculated from the partial pressures of K, Fe, Ni, Cu, S, and Cl output during calculations identical to those of Ebel and Alexander (2011), but including Cu. That is, reduction of partial pressures by condensation of metal, etc., were accounted for accurately. The affinity of djerfisherite of composition $X_2 = 0.0193$, $X_3 = 0.0631$ was calculated at 10 K intervals from 1600 to ~1100 K. Figure 5 illustrates the temperature, as a function of dust enrichment, at which djerfisherite becomes stable during cooling for systems of solar composition enriched up to 1000x in a C-IDP type dust.

**Discussion**

The calculated stability relations (Fig. 5) for djerfisherite closely correspond to petrographic observations of metal-sulfide and silicate chondrule paragenesis in enstatite chondrites (Fig. 4; El Goresy et al. 1988; Lin and El Goresy 2002). Oldhamite (CaS) is found in both chondrules and metal clasts, and is calculated to be stable but not abundant in the lower temperature range of silicate liquids. Djerfisherite is never found in chondrules, and is calculated to be stable only at temperatures below the liquid stability field. All Fe-Ni metal is calculated to condense from vapor at much higher temperature than does the djerfisherite that must form at the expense of metal alloy. Djerfisherite encloses numerous metal blebs, consistent with its formation by reaction of pre-condensed metal with vapor.

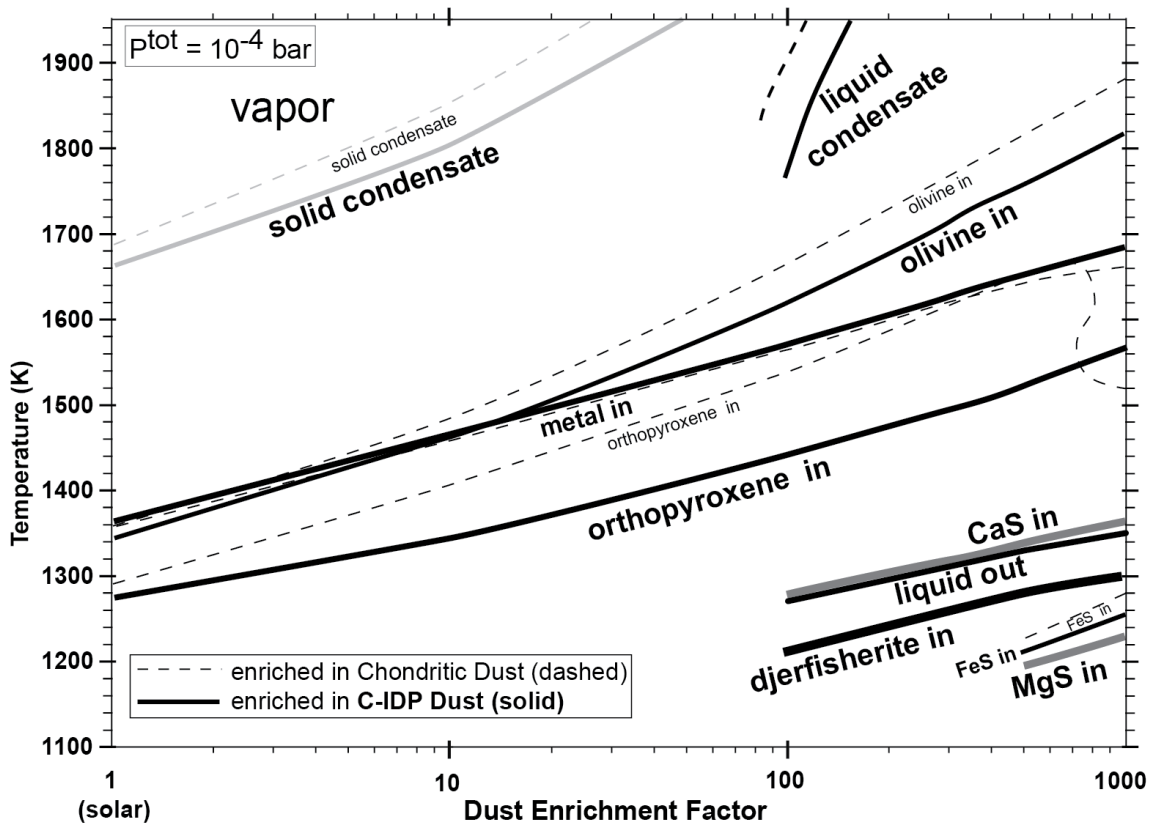

**Figure 5**: Djerfisherite stability calculated for C-IDP dust enriched systems at $P^{tot} = 10$Pa. Dashed lines show selected results of Ebel (2006, plate 10) for more oxidizing chondritic dust enriched systems. Phase boundaries for all phases except djerfisherite are as reported by Ebel and Alexander (2011).

Some details of the petrographic observations do not correspond well with results of the calculation. For example, breakdown of djerfisherite to troilite (FeS) is predicted. However, primary troilite, forming at temperatures above djerfisherite, is not predicted, most probably because the stabilizing effects of Cr and Ni in troilite are not currently considered (cf. Table 2). Similarly, the condensation of (Mg,Fe,Mn)S, niningerite (cf. Table 2) at higher temperatures than djerfisherite is also not preducted, because only the MgS endmember has been considered here. Nevertheless, the closeness in predicted condensation temperatures for the MgS niningerite endmember and djerfisherite (Fig. 5) makes it likely that explicit provisions for niningerite solid solution properties will remedy this situation.

Textures of coexisting djerfisherite and niningerite clearly demonstrate that the former condensed onto preexisting niningerite (El Goresy et al. 1988, their Figs 2, 13). While the K/Cu atom ratio in EH chondrites is ~7, the ratio in SAH 97097 djerfisherite (Table 2) is only ~2. The remaining K must reside elsewhere, but calculations do not indicate that chondrule silicate glass should contain substantial $K_2O$, and El Goresy et al. (1988) note the low K/Na ratio in chondrule mesostasis in EH3 chondrites. It may be that much of the primary djerfisherite has been removed by subsequent processing in most, if not all, of the enstatite chondrites, including least equilibrated EH3. The fates of Cu, Cl, and K, and the source of Zn in sphalerite, upon breakdown of djerfisherite in the Qingzhen reaction (El Goresy et al. 1988) are not known. Calculations of the stabilities of $NaCrS_2$ (caswellsilverite) and Ga-bearing sphalerite in these systems would also be interesting.

We have demonstrated that the K-rich sulfide djerfisherite is stable against evaporation in vapors highly enriched in a C-IDP type of nebular dust. Such environments may have characterized the formation region of enstatite chondrites (EC), and possibly the inner solar system feeding zone of Mercury. Indeed, links between EC or aubrites and Mercury include reduced (low-FeO) silicates (Robinson and Taylor 2001), high S/Si ratios (Nittler et al. 2011), and spectral similarity of the Mercurian surface to EC partial melts (McCoy et al. 1999; Burbine et al. 2002), and the possible presence of oldhamite (CaS) in surface rocks (Weider et al., 2012). Therefore, hypotheses for the formation of the EC and aubrite parent asteroid(s) are also relevant to the formation of the precursor materials of Mercury. While Mercury's K/Th ratio is similar to that of Mars and Earth (Peplowski et al. 2011), all the terrestrial planets are significantly depleted in K relative to the solar composition, consistent with Earth's mantle volatility trend (McDonough 2003). The present work demonstrates that the same nebular condensation conditions that stabilize Ca-, Mg-, and Fe-sulfides at high temperature also stabilize K-rich sulfide. High sulfur fugacities are likely to have significantly affected the fate of K (and Na) during Mercury's accretion, differentiation, and magma generation.

It is of note that in the EH3 chondrite Qingzhen, the primary, Ga-rich (i.e., > 1 wt%) sphalerites contain the highest FeS contents (to 0.58 $X_{FeS}$), whereas Ga-poor, secondary sphalerite produced during the Qingzhen reaction (QR, El Goresy et al. 1988; sph fragments enclosing schreibersite) have $0.44 < X_{FeS} < 0.5$ (Lin and El Goresy 2002, sph ~ 0.05 $X_{MnS}$). According to the calibration of Balabin and Urosov (1995) for sphalerites coexisting with troilite and Fe metal in the simple system Zn-Fe-S (c.f. Balabin and Sack 2000; Sack and Lichtner 2009), these FeS contents place maximum bounds on the temperatures of formation of primary and QR sphalerites of 1097 and > 473 K. A temperature of > 473 K for the breakdown of djerfisherite by QR is possibly below that inferred for the breakdown of djerfisherite by reaction (1) in the simple system K-Fe-S-Cl, 597 ± 28 K. As noted above, the Cu substitution in chondritic djerfisherites appears to stabilize them to temperatures well below 473 K, with this stabilization resulting from the strong preference of Cu for the B site, a result qualitatively confirmed by low temperature structure refinements (e.g.,

Czamanske et al. 1979; Evans and Clark 1981; Zaccarini et al. 2007). Exactly how strong this preference is, and the precise temperature of reaction (1), are in need of experimental clarification.

**Conclusions**

Djerfisherite is a primary condensate above 1000 K during cooling of solar composition vapors enriched > 500x in anhydrous chondritic interplanetary dust (C-IDP). Predicted petrogenetic relations are consistent with meteorite petrography. Elements S, K, and Cl are, therefore, more refractory in such systems than in more oxidized nebular chemical environments. These results provide a scenario for formation of the high-temperature metal-sulfide "chondrules" in enstatite chondrites, along with their coexisting silicate chondrules. If such materials were abundant in the inner solar system, they would be ubiquitous precursors of the innermost planets, explaining the S-, and K-rich nature of Mercury's exposed mantle. The potential role of djerfisherite in contributing K to planetary cores remains highly speculative, and its elucidation requires further research. The equation of state for djerfisherite presented here allows informed exploration of the thermodynamic conditions governing djerfisherite crystallization in terrestrial potassic igneous rocks.

**Acknowledgements**

This manuscript was greatly improved by reviews from Dr. A. El Goresy and an anonymous reviewer. This research has made use of the National Aeronautics and Space Administration's Astrophysics Data System Bibliographic Services. Research was supported by the American Museum of Natural History, National Aeronautics and Space Administration grant NNX10AI42G (DSE), and the friendly staff of OFM Research (ROS). Dr. Y. Lin kindly shared complete EPMA analyses for Qingzhen djerfisherites. We wish to dedicate this paper to Dr. Ahmed El Goresy upon his acceptance of the Leonard Medal of the Meteoritical Society in 2013, and to Dr. Ian Carmichael, whose interest in potassic igneous rocks was certainly not earthbound.

**References**

Anderson GM, Crerar D (1993) Thermodynamics in Geochemistry. Oxford U Press, New York.

Balabin AI, Sack RO (2000) Thermodynamics of (Zn,Fe) sphalerite: a CVM approach based on large basis clusters. Mineral Mag 64:943-963

Balabin AI, Urosov VS (1995) Recalibration of the sphalerite cosmobarometer: Experimental and theoretical treatment. Geochim Cosmochim Acta 59:1401-1410

Barkovi AY, Laajoki KVO, Gehör SA, Yakovlev YN, Taikina-Aho O (1997) Chlorine-poor analogues of djerfisherite – thalfenisite from Noril´sk, Siberia and Salmagorsky, Kola Peninsula, Russia. Can Mineral 35:1421-1430

Benz W, Slattery WL, Cameron AGW (1988) Collisional stripping of Mercury's mantle. Icarus 74:516-528

Benz W, Anic A, Horner J, Whitby JA (2007) The origin of Mercury. Space Sci Rev 132:189-202

Burbine TH, McCoy TJ, Nittler LR, Benedix GK, Cloutis EA, Dickinson TL (2002) Spectra of extremely reduced assemblages: Implications for Mercury. Meteor Planet Sci 37:1233-1244

Chabot NL, Drake MJ (1999) Crystallization of magmatic iron meteorites: the role of mixing in the molten core. Meteor Planet Sci 34:235–246

Chase MW Jr (1998) NIST-JANAF thermochemical tables, vol 9, 4th edn, Journal of Physical and Chemical Reference Data, National Institute of Standards and Technology, Washington DC

Clarke DB (1979) Synthesis of nickeloan djerfisherites and the origin of potassic sulfides at the Frank Smith Mine. In Boyd FR, Meyer HOA (eds) The Mantle Sample: Inclusion in Kimberlites and Other Volcanics, Proc. Second Intl. Kimberlite Conf., Am Geophys Union, Washington D.C., pp 300-30

Clarke DB, Mitchell RH, Chapman CAT, MacKay RM (1994) Occurrence and origin of djerfisherite from the Elwin Bay kimberlite, Somerset Island, Northwest Territories. Can Mineral 32:815-823

Clay PL, King A, Wieler R, Busemann H (2012) Noble gas chronology of EH5 chondrite St. Mark's - an in-vacuo etch experiment. Meteor Planet Sci Suppl 47:5335

Craig JR, Barton PB Jr (1973) Thermochemical approximations for sulfosalts. Econ Geol 68:493-506

Czamanske GK, Erd RC, Sokolova NM, Dobrovol'skaya MG, Dmitrievea MT (1979) New data on rasvumite and djerfisherite. Am Mineral 64:776-778

Desch SJ, Connolly HC Jr (2002) A model of the thermal processing of particles in solar nebula shocks: Application to the cooling rates of chondrules. Meteor Planet Sci 37: 183-207

Dmitrievea MT, Hyukhin VV (1975) Crystal structure of djerfisherite. Doklady Akademiia Nauk USSR 223:343-346 (transl. Soviet Physics Doklady 20:469-470, 1976).

Ebel DS (2006) Condensation of rocky material in astrophysical environments. In: Lauretta DS, McSween H (eds), Meteorites and the Early Solar System II, University of Arizona, Tucson, pp 253-277

Ebel DS, Alexander CMO'D (2011) Equilibrium condensation from chondritic porous IDP enriched vapor: Implications for Mercury and enstatite chondrite origins. *Planet Space Sci* 59:1888-1894

Ebel DS, Grossman L (2000) Condensation in dust-enriched systems. Geochim Cosmochim Acta 64:339-366

Ebel DS, Sack RO (1994) Experimental determination of the free energy of formation of freibergite fahlore. Geochim Cosmochim Acta 58:1237-1242

El Goresy A, Grögler N, Ottemann J (1971) Djerfisherite composition in Bishopville, Peña Blanca Springs, St. Marks and Toluca Meteorites. Chem Erde 30:77-82

El Goresy A, Hideo Y, Ehlers K, Woolum D, Pernicka E (1988) Qingzhen and Yamato-691: A tentative alphabet for the EH Chondrites. Proc. NIPR Symp Antarct Meteorites 1:65-101

Evans HT Jr, Clark JR (1981) The crystal structure of bartonite, a potassium iron sulfide, and its relationship to pentlandite and djerfisherite. Am Mineral 66:376-384

Fegley B Jr, Cameron AGW (1987) A vaporization model for iron/silicate fractionation in the Mercury protoplanet. Earth Planet Sci Lett 82:207-222

Fuchs LH (1966) Djerfisherite, alkali-copper sulfide: a new mineral from enstatite chondrites. Science 153:166-167

Goettel KA (1976) Models for the origin and composition of the earth, and the hypothesis of potassium in the Earth's core. Surveys in Geophysics 2:369-397

Goettel KA (1988) Present bounds on the bulk composition of Mercury: Implications for planetary formation processes. In: Vilas F, Chapman CR, Matthews MS (Eds.) Mercury, University of Arizona, Tucson, pp 613-621

Henderson CMB, Kogarko LN, Plant DA (1999) Extreme closed system fractionation of volatile-rich, ultrabasic peralkaline melt inclusions and the occurrence of djerfisherite in the Kugda alkaline complex, Siberia. Mineral Mag 63:433-438

Javoy M (1995) The integral enstatite chondrite model of the Earth. Geophys Res Lett 22:2219–2222

Keil K (1968) Mineralogical and chemical relationships among enstatite chondrites. J Geophys Res 73:6945-6976

Keil K (1989) Enstatite meteorites and their parent bodies. Meteoritics 24: 195-208.

Keil K (2010) Enstatite achondrite meteorites (aubrites) and the histories of their astereoidal parent bodies. Chem Erde 70:395-317

Kimura M (1988) Origin of opaque minerals in an unequilibrated enstatite chondrite, Yamato-691. Proc. NIPR Symp Antarctic Meteorites 1:51-64

Kracher A, Kurat G, Buchwald VF (1977) Cape York; The extraordinary mineralogy of an ordinary iron meteorite and its implication for the genesis of IIIAB irons. Geochem J 11:207-217.

Lehner SW, Buseck PR, McDonough WF (2010) Origin of kamacite, schreibersite, and perryite in metal-sulfide nodules of the enstatite chondrite Sahara 97072 (EH3). Meteor Planet Sci 45:289-303

Lewis JS (1971) Consequences of the presence of sulfur in the core of the Earth. Earth Planet Sci Lett 11:130-134.

Lin Y, El Goresy A (2002) A comparative study of opaque phases in Qingzhen (EH3) and MacAlpine Hills 88136: Representatives of EH and EL parent bodies. Meteor Planet Sci 37:577-599

Lodders K (1995) Alkali elements in the Earth's core: Evidence from enstatite meteorites. Meteor Planet Sci 30:93-101

Lodders K (2003) Solar system abundances and condensation temperatures of the elements. Astrophys J 591:1220-1247

Lodders K, Fegley B Jr (1998) The Planetary Scientist's Companion. Oxford U Press, New York

McCoy TJ, Dickinson TL, Lofgren GE (1999) Partial melting of the Indarch (EH4) meteorite: A textural, chemical and phase relations view of melting and melt migration. Meteor Planet Sci 34:735-746

McDonough WF (2003) Compositional model for the Earth's core. In: Carlson RW (ed) The Mantle and Core, pp 547-568, In: Holland HD, Turekian KK (eds) Treatise on Geochemistry, Elsevier, vol 2

McDonough WF, Sun S-S (1995) The composition of the Earth. Chem Geol 120:223– 253

McNally C, Hubbard A, Mac Low M-M, Ebel DS, and D'Alessio P (2013) Mineral processing by short-circuits in protoplanetary disks. Astrophys J (in press; http://arxiv.org/pdf/1301.1698.pdf).

Nittler LR, Starr RD, Weider SZ, McCoy TJ, Boynton WV, Ebel DS, Ernst CM, Evans LG, Goldsten JO, Hamara DK, Lawrence DJ, McNutt RL Jr, Schlemm CE II, Solomon SC, Sprague AL (2011) The major-element composition of Mercury's surface from MESSENGER x-ray spectrometry.  Science 333:1847-1850

Peplowski PN, Evans LG, Hauck SA II, McCoy TJ, Boynton WV, Gillis-Davis J, Ebel DS, Goldsten JO, Hamara DK, Lawrence DJ, McNutt RL Jr, Nittler LR, Solomon SC, Rhodes EA, Sprague AL, Starr RD, Stockstill-Cahill KR (2011) Radioactive elements on Mercury's surface from MESSENGER: Implications for the planet's formation and evolution. Science 333:1850-1852

Rajamani V, Prewitt CT (1973) Crystal chemistry of natural pentlandites. Can Mineral 12:178-187

Rama Murthy V, van Westrenen W, Fei Y (2003) Experimental evidence that potassium is a substantial radioactive heat source in planetary cores. Nature 423:163-165

Rietmeijer FJM (2002) The earliest chemical dust evolution in the solar nebula. Chem Erde 62:1-45

Robinson MS, Taylor GJ (2001) Ferrous oxide in Mercury's crust and mantle. Meteor Planet Sci 36:841-847

Sack RO, Ebel DS (2006) Thermochemistry of sulfide mineral solutions. Rev Mineral Geochem 61:265-364

Sack RO, Lichtner PC (2009) Constraining compositions of hydrothermal fluids in equilibrium with polymetallic ore forming sulfide assemblages. Econ Geol 104:1249-1264

Sack RO, Ebel DS, O'Leary MJ (1987) Tennahedrite thermochemistry and metal zoning. In Chemical Transport in Metasomatic Processes. Helgeson HC (ed) D. Reidel, Dordrecht, Holland, pp701-731

Sharygin VV, Golovin AV, Pokhilenko NP, Kamenetsky WS (2007) Djerfisherite Udachnaya-East pipe kimberlites (Sakha-Yakutia, Russia): paragenesis, composition and origin. Eur J Mineral 19:51-63

Sommerville M, Ahrens TJ (1980) Shock compression of $KFeS_2$ and the question of potassium in the core. J Geophys Res 85:7016-7024

Tani BS (1977) X-ray study of $K_6LiFe_{24}S_{26}Cl$, a djerfisherite-like compound. Am Mineral 62:819-923

Thompson JB Jr (1969) Chemical reactions in crystals. Am Mineral 54:341-375

Urey HC (1955) The cosmic abundances of potassium, uranium, and the heat balance of the earth, the moon, and Mars. Proc Nat Acad Sci 41:127-144

van Acken D, Humayun M, Brandon AD, Peslier AH (2012) Siderophile trace elements in metals and sulfides in enstatite achondrites record planetary differentiation in an enstatite chondritic parent body. Geochim Cosmochim Acta 83:272-291.

van Westrenen W, Rama Murthy V (2006) Bulk Earth compositional models are consistent with the presence of potassium in Earth's core. Workshop Early Planet Differentiation, LPI. p113-114
http://www.lpi.usra.edu/lpi/contribution_docs/LPI-001335.pdf

Wasserburg GJ, MacDonald GJF, Hoyle F, Fowler WA (1964) Relative contributions of Uranium, Thorium, and Potassium to heat production in the Earth. Science 143:465–467

Wasson JT, Kallemeyn GW (1988) Composition of chondrites. Phil Trans R Soc London A 325:535-544

Watters TR, Prinz M (1979) Aubrites: Their origin and relationship to enstatite chondrites. Proc Lunar Planet Sci Conf 10:1073-1093

Weidenschilling SJ (1978) Iron/silicate fractionation and the origin of Mercury. Icarus 35:99-111

Weider SZ, Nittler LR, Starr RD, McCoy TJ, Stockstill-Cahill KR, Byrne PK, Denevi BW, Head JW, Solomon SC (2012) Chemical heterogeneity on Mercury's surface revealed by the MESSENGER x-ray spectrometer. J Geophys Res 117: E00L05

Weisberg MK, Kimura M (2012) The unequilibrated enstatite chondrites. Chem Erde 72:101-115

Zaccarini F, Thalhammer OAR, Princivalle F, Lenaz D, Stanley CJ, Garuti G (2007) Djerfisherite in the Guli dunite complex, polar Siberia: a primary or metasomatic phase? Can Mineral 45:1201-1221

**Appendix I: Ordering Calculation**

$(1 - X_2 - X_3 - \frac{24}{25}(s_1 + s_2))(X_2 - \frac{1}{25}s_1) = \exp(\frac{25}{48}\Delta\bar{G}_{Ni-Fe}^{o-ORD}/RT)(1 - X_2 - X_3 + \frac{1}{25}(s_1 + s_2))(X_2 + \frac{24}{25}s_1)$

defining $K_1 \equiv \exp(\frac{25}{48}\Delta\bar{G}_{Ni-Fe}^{o-ORD}/RT)$, and $D \equiv X_2 - (X_2)^2 - X_2 X_3 + \frac{1}{25} X_2 s_2$,

and $E \equiv \frac{24}{25}(1 - X_3 + \frac{1}{25}s_2) - \frac{23}{25}X_2$, and $F \equiv \frac{24}{625}$, and $X \equiv X_2 - (X_2)^2 - X_2 X_3 - \frac{24}{25}X_2 s_2$, and $Y \equiv \frac{1}{25}(-1 - 23 X_2 + X_3 + \frac{24}{25}s_2)$, and $Z \equiv \frac{24}{625}$

we have $0 = (K_1 D - X) + (K_1 E - Y)s_1 + (K_1 F - Z)(s_1)^2$.

Substituting $X_2$ for $X_3$, and $X_3$ for $X_2$, and $s_1$ for $s_2$, and $s_2$ for $s_1$, and $\Delta\bar{G}_{Cu-Fe}^{o-ORD}$ for $\Delta\bar{G}_{Ni-Fe}^{o-ORD}$ in the first expression above, leads to an analogous quadratic equation in $s_2$ in similarly substituted variables.

**Appendix II: EPMA Method**

Electron probe microanalysis was performed with one micron spot size, at 15 keV accelerating voltage and 20 nA beam current. Natural and synthetic standards were used to calibrate on the Kα lines for 16 elements using wavelength dispersive spectrometers (WDS), and one element using energy dispersive analysis (EDS). Unknowns were analyzed in five passes as listed in Table A-1, and an EDS spectrum was collect ed on each spot with a 240 sec dwell time.

| element | xtal | pk time (s) | standard | Det.Lim ppm | pass |
|---------|------|-------------|----------|-------------|------|
| Na | TAP | 60 | jadeite | 213 | 1 |
| Si | TAP | 60 | Si metal | 152 | 2 |
| Cr | LLiF | 30 | Cr metal | 414 | 1 |
| Co | LLiF | 30 | Co metal | 0 | 2 |
| Ni | LLiF | 45 | Ni metal | 484 | 3 |
| Cu | LLiF | 45 | Cu metal | 1223 | 4 |
| Ti | LiF | 30 | rutile | 615 | 1 |
| Mn | LiF | 60 | Mn metal | 479 | 2 |
| Fe | LiF | 60 | Fe metal | 575 | 3 |
| Mg | TAP | 120 | MgCr2O4 | 102 | 1 |
| Al | TAP | 60 | MgAl2O4 | 147 | 2 |
| Cl | PET | 30 | scapolite | 355 | 1 |
| K | PET | 30 | orthoclase | 263 | 2 |
| P | PET | 30 | schreibersite | 277 | 3 |
| S | PET | 30 | troilite | 469 | 4 |
| Ca | PET | 30 | apatite | 0 | 5 |
| Zn | EDS | 240 | sphalerite | 12 | all |

**Table A-1:** EPMA conditions. Dwell time on peak (pk time) is twice the dwell time on background wavelengths. Detection limits (ppm) are those recorded by the CAMECA peak-site software in the output file.